\documentclass[paper]{JHEP}
\usepackage{amsmath}
\usepackage{amssymb}
\usepackage{epsfig}
\usepackage{cite}

\def\as{\alpha_{\mbox{\scriptsize s}}}
\def\bas{{\bar\alpha}_s}
\def\qq{q\bar{q}}
\def\ee{e^+e^-}

\def\Cout{C_{\rm out}}
\def\out{{\rm out}}

\def\Eout{E_{\rm out}}
\def\tout{\theta_{\rm out}}

\def\nuout{\nu_{\rm out}}

\def\cR{{\cal R}}
\def\Cin{C_{\rm in}}
\def\Ein{V}
\def\tin{\theta_{\rm in}}

\def\tot{{\rm tot}}

\def\Ctot{C_{\rm tot}}

\def\De{\Delta}
\def\Om{\Omega}
\def\om{\omega}

\def\lam{\lambda}

\def\lqcd{\Lambda_{\mbox{\scriptsize QCD}}}

\def\cO#1{{\cal{O}}\left(#1\right)}

 \def\caja{\mathsurround=0pt}
 \def\eqalign#1{\,\vcenter{\openup1\jot \caja \ialign{\strut
 \hfil$\displaystyle{##}$&$ \displaystyle{{}##}$\hfil\crcr#1\crcr}}\,}

\def\ng{\rm ng}

\def\ee{e^+e^-}

\title{On 
  large  angle multiple gluon radiation }

\author{Yu.L. Dokshitzer $^{{a}}$\footnote{On leave from St. Petersburg  
Nuclear Institute, Gatchina, St. Petersburg 188350, Russia}
{ }and G. Marchesini $^{{a,b}}$
\\
$^{\mbox{a}}$ LPTHE, Universit\'es Paris 6 et 7, 
CNRS UMR 7589,
Paris, France\\ $^{\mbox{b}}$ Dipartimento di Fisica,
Universit\`a di Milano-Bicocca and INFN, Sezione di Milano, Italy}

\abstract{ 
 Jet shape observables which involve measurements restricted to a {\em
 part}\/ of phase space are sensitive to multiplication of soft gluon
 with large relative angles and give rise to specific single
 logarithmically enhanced (SL) terms (non-global logs).
 
 We consider {\em associated}\/ distributions in two variables which
 combine measurement of a jet shape $V$ in the whole phase space
 (global) and that of the transverse energy flow away from the jet
 direction, $\Eout$ (non-global).

 We show that associated distributions factorize into the global
 distribution in $V$ and a factor that takes into account
 SL contributions from multi-gluon ``hedgehog'' configurations in all
 orders. The latter is the same that describes the single-variable
 $\Eout$ distribution, but evaluated at a rescaled energy $VQ$.  }


\preprint{Bicocca--FT--03--6\\
     LPTHE-03-10  \\
     hep-ph/0303101\\
     February 2003}

\begin{document}

\setcounter{footnote}{0}

\section{Introduction}

 In a theory with dimensionless coupling one would naively expect
 cross sections to scale in a simple manner at very large $Q^2$,
 $Q^2\gg m^2$, with $m$ a generic mass scale.  However, in QCD this is
 not true for two reasons.  Firstly, due to UV divergences, the
 effective interaction strength does vary with scale (as in any
 quantum field theory with dimensionless coupling). As a result
 perturbative (PT) corrections to cross sections slowly decrease with
 $Q^2$ as powers of $\as\propto 1/\ln (Q^2/\lqcd^2)$.  Secondly, there
 are, generally speaking, collinear and infrared divergences
 (collinear 2-parton splittings, soft gluon radiation).  As well known
 (Bloch-Nordsieck~\cite{BN}), in ``good'' inclusive observables which
 do not include observation (fixing the momentum) of a single hadron,
 either in the initial or final state, the logarithmic collinear and
 infra-red divergences cancel.

 The classical examples are the total cross section of $e^+e^-$
 annihilation into hadrons, $\tau\to$ hadrons decay width.  Here the
 collinear and infra-red divergences, present in both real and virtual
 corrections, cancel completely in the (unrestricted) sum. Since here
 we don't have dimensional parameters other than $Q^2$ (in the
 $m^2/Q^2\to 0$ limit), the total cross section is given by the simple
 Born expression modulo calculable $\as^n(Q^2)$ corrections.
$$ 
\sigma(Q) = \sigma^{\mbox{\scriptsize Born}}(Q) \cdot
 g\left(\as(Q^2)\right), \quad g(0)=1 \qquad (Q^2\gg m^2).  
$$

 Moving to {\em less inclusive}\/ measurements one faces different
 situations.  The first case involves fixing (measuring) momentum of a
 hadron, e.g.\ that of the initial proton in DIS, DY, \ldots
 (structure functions, SF) or of a final hadron (fragmentation
 function). Then soft divergences still cancel but collinear ones do
 not, making such observables not calculable at the parton level.
 These effects, however, turn out to be universal and, given a proper
 technical treatment, can be {\em factored out}\/ as non-perturbative
 (NP) inputs. What remains under control then is only the
 $Q^2$-dependence (scaling violation pattern).  

 Collinear divergences in the {\em final}\/ state may be avoided
 altogether if one looks at {\em energy flows}\/ rather than
 individual hadrons.  More generally, one can study a family of the
 so-called collinear and infra-red safe (CIS) jet shapes $V$,
\begin{equation}
 \label{eq:Vdef} V = \sum_i v(k_i), 
\end{equation} 
 where the sum runs over all particles (hadrons) in the final state,
 $v(k)$ being a contribution of a single particle with 4-momentum
 $k$. Being (by construction) {\em linear}\/ in particle momenta, such
 observables are also free from collinear (and soft) {\em
 divergences}.

 However, here the cancellation of real and virtual effects is not
 complete and leaves trace in the PT-calculable distributions over
 $V$. Indeed, taking the value of a generic jet shape observable $V\ll
 1$ we squeeze the phase space thus inhibiting {\em real}\/ parton
 production and multiplication. Since the {\em virtual}\/ PT radiative
 contributions remain unrestricted, the {\em divergences}\/ do cancel but
 produce finite but logarithmically enhanced leftovers:
\begin{equation} 
 \label{eq:Sigmadef} 
 {\Sigma(Q,V)} \equiv \int_0^V dV \frac{d\sigma(V)}{\sigma_\tot\, dV} 
 = f\left(\as(Q^2), \ln V\right) .  
\end{equation}

 Each gluon emission brings in at most two logarithms (one of
 collinear, another of infra-red origin).  These leading contributions
 are due to multiple soft gluon bremsstrahlung off the primary hard
 partons that form the underlying event, which can be looked upon as
 independent gluon radiation and can be easily treated.
 A clever reshuffling of PT series, based on universal nature of soft
 and collinear radiation (factorization) results in 
 the {\em exponentiated}\/ answer in the form
\begin{equation}
\label{eq:exp}
  \ln \Sigma(Q,V) = \sum_{n=1}^\infty \as^n(Q^2)\left( A_n\ln^{n+1}V 
                + B_n\ln^n V + \cdots\right) . 
\end{equation}
In what follows we will discuss only these two series of term and
neglect subleading small corrections of the order of $\as$, $\as^2\ln
V$, \ldots. The $A_n$ series is referred to as double logarithmic (DL)
and $B_n$ as single logarithmic (SL).

The whole leading series $A_n$ actually originates from the effect of
running of the coupling in the basic one gluon radiation term,
$\as\ln^2V$, $n=1$.  Indeed, the expansion in $\as(Q^2)$ is pretty
artificial since in reality it is a wide range of scales, $(VQ)^2\ll
k^2 \ll Q^2$, at which the coupling $\as(k^2)$ actually enters:
\begin{equation}
 A_1 \as(Q^2)\ln^2V \>\Longrightarrow\> A_1 \int_{(VQ)^2}^{Q^2}
 \frac{dk_t^2}{k_t^2} \,\as(k_t^2) \ln\frac{Q}{k_t} =
 \sum_{n=1}^\infty A_n\cdot\as^n(Q^2) \ln^{n+1}V,
\end{equation} 
where $k_t$ is the gluon transverse momentum and the logarithmic
factor is due to the soft enhancement.  As a result, the coefficients
$A_n$ are straightforward to obtain.  It is important to realise that
what makes the coupling run in Minkowskian observables is collinear
gluon splittings (into gluons with energies of the same order, ``hard
splitting'', and $\qq$ pairs) and the inclusive CIS nature of the
observable (see, e.g.\ ~\cite{WDM}).
   
Let us remark in passing that the scales {\em below}\/ $(VQ)^2$ with
$k_t^2$ running down to zero are actually present as well. They,
however, do not contribute at the PT level but give rise to the NP
power suppressed corrections (see~\cite{WDM,NP0}).

The subleading series $B_n$ start from the first SL correction $\as\ln
V$ ($n=1$) (of either collinear or soft origin).  In higher orders,
$n\ge2$, taking care of SL terms involves a careful treatment of $\as$
(physical scheme), of its running argument, as well as precise fixing
of the scales of the leading logarithmic terms in \eqref{eq:exp}.

In the case of the so-called ``global'' observables, that is
measurements in which the observable \eqref{eq:Vdef} accumulates
contributions from final state particles in the whole phase space, all
SL contributions are generated by a careful treatment of gluon
bremsstrahlung off the primary partons (Sudakov exponentiation).
Obviously, there exists also gluon multiplication and, in particular,
gluon bremsstrahlung off {\em secondary}\/ partons.  As already
stated, {\em collinear}\/ ``hard'' gluon decays make the coupling run.
As for {\em soft}\/ gluon bremsstrahlung off secondary gluons, it does
not affect global observables at the DL+SL accuracy
\eqref{eq:exp} (the first correction being $\cO{\as^2\ln V}$).
Global observables have been intensively studied in the literature
both perturbatively~\cite{PTcollect,Broad} (DL+SL as in
\eqref{eq:exp}) and with account of the first leading NP power
correction $\cO{1/Q}$~\cite{NP0}.
   
However, as was recently noticed by Dasgupta and Salam, certain jet
shape measurements turn out to be sensitive to soft gluon
multiplication effects at the SL level starting from $n=2$.  These
observables involve measurements restricted to a {\em part}\/ of phase
space and were correspondingly dubbed {\em non-global}~\cite{DS}.  The
simplest example is given by particle transverse energy flow $\Eout$ in
$\ee$ annihilation, measured in an angular (pseudorapidity) region
$\Cout$ which is away from the jet direction (thrust axis) by a finite
angle $\theta_0$:
\begin{equation}
\label{eq:Soutdef}
 \Sigma_\out(Q,\Eout), \quad \Eout = \sum_{i\in\Cout} k_{ti}.
\end{equation}
It is a CIS observable. Moreover, there is obviously no collinear
enhancement ($A_n\equiv 0$).  The leading effect is SL --
exponentiation of independent large angle soft gluon emission off the
primary jets. Apart from it, however, additional contributions of the
same order emerge here due to soft gluon-gluon correlations, which
were absent in global observables.
  
Imagine a system of gluons whose energies are strongly ordered:
$$
k_\ell \ll k_{\ell-1} \ll \ldots \ll k_1 \ll p \simeq Q/2.
$$ 
Then only the {\em hardest}\/ of the gluons belonging to $\Cout$ would
contribute to the observable while the contributions of all other
(much softer) gluons would cancel against corresponding virtual
corrections.  This means that it suffices to consider multi-gluon
systems with the {\em softest}\/ offspring $k_n\in \Cout$ being the
only one to be measured, while all harder companions do not contribute
to the observable, $k_i\in\Cin$, $i<n$, where $\Cin$ is the
complementary angular region, close to the jets.
 
Such configurations of $n\ge2$ gluons contribute to the integrated
distribution \eqref{eq:Soutdef} at the SL level as
$(\as\ln(Q/\Eout))^n$.  This is the origin of the Dasgupta--Salam
discovery.

These specific subleading contributions are not easy to analyse
analytically order by order. In spite of the fact that in the strong
energy ordered kinematics all $(n+2)!$ amplitudes are known and given
by soft insertion rules, to give a compact answer to the {\em
square}\/ of the $n$-gluon matrix element is possible only in the
large-$N_c$ limit (see~\cite{BCM}).  Moreover, angles between gluons
are of the order one (hedgehog configurations), so that the phase
space integrations can be handled only numerically.  As we shall show
below, the all-order resummation of these contributions can be carried
out in the large-$N_c$ limit. This leads to an evolution
equation~\cite{BMS} which has a highly non-linear
structure. Therefore, its solution can be given in an analytic form
only in an academic high-$Q$ limit ($\as\ln(Q/\Eout)\gg 1$).

Recall that non-linear evolution equations for generating functionals
describing multi-parton ensembles have a long history. In the leading
{\em collinear}\/ approximation they, in particular, form the basis
for Monte Carlo generators and for the standard theoretical jet
studies (jet rates, in-jet particle multiplicities and spectra,
etc.)~\cite{Darwin,BCM}. When large angle soft gluon radiation becomes
important (as, e.g., in inter-jet particle flows -- the bread for the
so-called string effect), in the {\em collinear}\/ approximation (that
is when other secondary gluons stay quasi-parallel to the primary hard
parton) the QCD coherence is at work. As a result, a single soft gluon
emission at large angles is very simple and is determined by the total
colour charge of the jet.

 However, when the large-angle hedgehogs are considered, the new
 evolution equation for such systems has an essentially different (and
 more complicated) structure which, as mentioned above, can be
 practically handled only in the large-$N_c$ approximation.
 
It is important to stress that most of the actual experimental
measurements are subject to non-global effects.  Experiments often (if
not always) involve phase space restriction. For example,
\begin{itemize}
\item the very first CIS jet cross section invented by Sterman and
  Weinberg back in 1979~\cite{SW} provides a perfect example of a
  non-global observable;
\item in hadron--hadron interactions accompanying hadron production is
  studied in a limited rapidity range (in particular, the famous {\em
    pedestal}\/ distributions in hard hadron--hadron collisions);
\item direct photon studies necessarily involve photon isolation
  criteria;
\item a family of QCD string/drag effects deals with particle flows in
  restricted inter-jet angular regions;
\item profiles of a separate jet (rather than studying shape
variables of the whole event). For example, characteristics of the
current fragmentation jet in DIS which is based in a one-hemisphere
particle selection.
\end{itemize}
All these and many other similar measurements contain non-global PT
corrections.  With exception of the current jet in DIS
(see~\cite{DSDIS}) and the case study of the $\Eout$ distribution in
$\ee$\cite{DScase,BMS,AS}, the non-global effects have not been
studied theoretically even at level of the first $(\as\log)^2$
correction.

In spite of being formally subleading, these effects may significantly
modify the PT predictions for the non-global observables, as was shown
by the case study of the $\Eout$ distribution in $\ee$ by Dasgupta and
Salam~\cite{DScase}.

Berger, Kucs and Sterman~\cite{BKS} have recently formulated a
programme of how to avoid non-global logarithms in a measurement of
transverse energy flow away from jets ($\Eout$).  They suggested to
squeeze the jets and thus suppress multi-gluon hedgehogs in the $\Cin$
region.  In $\ee$ this amounts to introducing the {\em associated}\/
distribution in two variables, $\Eout$ and $V$ (for example, $V=1-T$
with $T$ the thrust), and considering the region $V\ll1$ which selects
2-jet-like configurations,
$$
\Sigma_{2\ng}(Q,V,\Eout).
$$
They treated the region in which the two characteristic
scales of the problem are comparable, $VQ/\Eout=\cO{1}$, so that 
\begin{equation}
\label{eq:ratto}
  \as\ln\frac{VQ}{\Eout} 
\end{equation}
amounts to a negligible correction $\cO{\as}$.  The authors stated in
the Conclusions to~\cite{BKS} that in this approximation their
``formalism is sensitive mainly to radiation stemming directly from
primary hard scattering''.

In this paper we consider a general case\footnote{We take $V$ to be a
global observable, though one can equally well restrict $V$ to $\Cin$
as was done in~\cite{BKS}.}  of $\Eout$ being potentially much smaller
than $VQ$. We show that, having extracted the global DL and SL
enhanced term in $\ln V$, the remaining SL corrections $(\as\ln V)^n$
and $(\as\ln(Q/\Eout))^n$ conspire to produce the powers in
\eqref{eq:ratto}.  

We analyse these non-global logarithms in all orders and demonstrate
that the associated distributions factorize as
\begin{equation}
\label{eq:result0}
\Sigma_{2\ng}(Q,V,\Eout) = \Sigma(Q,V) \cdot \Sigma_{\out}(VQ,\Eout).
\end{equation}
Here $\Sigma(Q,V)$ is the standard global distribution
\eqref{eq:Sigmadef} and $\Sigma_{\out}(Q,\Eout)$ is the SL
distribution in $\Eout$ (at the same total energy $Q$) which takes
into account contributions from multi-gluon hedgehog configurations in
all orders and contains the full dependence on the geometry of the
measurement ($\theta_0$).
This is the main result of the paper.

We will also discuss the asymptotic behaviour which according to
\eqref{eq:result0} reduces to that of the simplest non-global
distribution $\Sigma_\out$ which had been studied in~\cite{BMS}.

Let us stress that one needs to keep under the best possible control
effects induced by radiation of relatively soft gluons not merely for
the sake of improving PT predictions.  What makes such studies even
more important and interesting is the fact that the physics of small
transverse momentum gluons is that of confinement.\footnote{In market
terms, a \$ $\!\!10^6$ problem~\cite{1000000}}

\section{Observables}

In this paper we will consider $e^+e^-$ annihilation into hadrons and
define a cone around the thrust axis and denote by $\Cout$ and $\Cin$
the regions outside and inside the solid cone (inter and intra jet
regions).

We start by considering the first factor in \eqref{eq:result0} that is
the (integrated) {\em global}\/ distribution $\Sigma(Q,V)$
defined as
\begin{equation}
\label{eq:tot}
 \Sigma(Q,V) = \sum_n\int \frac{d\sigma_n}{\sigma_\tot}
 \Theta\left(V - \sum_{i\in \Ctot}v(k_{i})\right)\,,
\end{equation}
with the sum extended to the full phase space $\Ctot=\Cin + \Cout$.
Here $d\sigma_n$ and $\sigma_\tot$ are the production and total cross
sections.
The probing function $v(k)$, linear in the particle momentum $k$, can
be represented as
\begin{equation}
\label{eq:v}
  v(k)=\frac{k_t}{Q}\cdot h(\eta)\,,
\end{equation}
with $k_t$ the transverse momentum with respect to the thrust axis
and $\eta$ the pseudorapidity. 
To get the distribution in thrust, $T$, we have to take
$h(\eta)=e^{-|\eta|}$ and the set $V = 1\!-\!T$. For small $1-T$,
summing over final particles we may neglect contributions from the
primary hard partons (quarks) since $h(\eta)$ vanishes at large
$|\eta|$, and restrict ourselves to to considering only secondary soft
gluons.
 
The distribution in broadening $B$, for example, corresponds to
$h(\eta)=1$, $V = 2B$. In this case, even for $B\ll1$, one needs to
include the contributions from the recoiling primary quarks.

Now we define analogously the {\em non-global}\/ distribution (the
second factor on the r.h.s.\ of \eqref{eq:result0}):
\begin{subequations}
\label{eq:2distr}
  \begin{equation}
    \label{eq:out}
 \Sigma_{\out}(Q,\Eout)  =  \sum_n\int \frac{d\sigma_n}{\sigma_\tot}
 \Theta\left(\Eout - \sum_{i\in \Cout} k_{ti}\right)\,,    
  \end{equation}
where we have chosen to measure the transverse energy flow accumulated in the
angular region $\Cout$ away from the jets.

Finally, as was suggested by Berger, Kucs and Sterman~\cite{BKS}, we
introduce the shape observable distribution in two variables (the {\em
  associated}\/ distribution)
\begin{equation}
  \label{eq:ng}
 \Sigma_{2\ng}(Q,V,\Eout) = \sum_n\int \frac{d\sigma_n}{\sigma_\tot}
 \> \Theta\left(V-\sum_{i\in \Ctot}v(k_{i})\right) \cdot 
 \Theta\left(\Eout-\sum_{i\in \Cout}k_{ti}\right). \quad { }  
\end{equation}
\end{subequations}

\section{Analysis and resummation}

We are interested in the soft region
\begin{equation}
\label{eq:soft}
   \Ein,\Eout\ll Q\,.
\end{equation}

\subsection{Mellin factorization}
First we recall how the resummation with the SL accuracy is done for
global observables.  To this end one has to factorize the
theta-function in \eqref{eq:tot} by the Mellin transform:
\begin{equation}
\label{eq:Mellintot}
\begin{split}
 &\Sigma(Q,V)\!=\! \int\frac{d\nu}{2\pi i \nu}\> e^{\nu \, V}\>
 \tilde\Sigma(Q,\nu)\,,\\ &\tilde\Sigma(Q,\nu) = \sum_n\int
 \frac{d\sigma_n}{\sigma_\tot}
\prod_{i\in \Ctot}\!\! u(k_i)\,, \quad u(k)=e^{-\nu\,v(k)}\,.
\end{split}
\end{equation}
As has been mentioned above, in the thrust case the contributions of
the primary partons can be dropped (since $v(k)$ vanishes at large
$|\eta|$).

In the case of broadening the recoiling primary partons do contribute.
This causes certain complication. This recoil effects can be taken
care of by expressing the momenta of recoiling quarks in terms of
those of the secondary partons via 3-momentum conservation. Thus, also
in the case of broadening one can restrict the product in
\eqref{eq:Mellintot} to the secondary soft partons, provided one
factorizes also the momentum conservation constraints by the Fourier
transform.  This amounts to incorporating into $\tilde \Sigma$ an
additional Fourier variable associated with transverse momentum
conservation.  Peculiarities of $B$ as a variable employed to squeeze
the jets, can be analyzed along the lines of~\cite{Broad}. This leads
to the known modification of the {\em global}\/ distribution but will
not affect the dependence on $\Eout$ in \eqref{eq:result0}.

In what follows to simplify the discussion we neglect this
complication and concentrate on the thrust case.

For the out-distribution we have
\begin{equation}
\label{eq:Mellinout}
\begin{split}
 \Sigma_\out(Q,\Eout) &= \int\frac{d\nuout}{2\pi i \nuout}\> e^{\nuout
 \, \Eout/Q} \> \tilde\Sigma_\out(Q,\nuout)\,,\\
 \tilde\Sigma_\out(Q,\nuout) &= \sum_n\int
 \frac{d\sigma_n}{\sigma_\tot} \prod_{i\in \Ctot}\!\! u_\out(k_i)\,,
 \\ u_\out(k) &= \tin + \tout e^{-\nuout\,k_t/Q} = 1-\tout
 \left(1-e^{-\nuout\,k_t/Q}\right) .
\end{split}
\end{equation}
Here $\tin$, $\tout$ are the support functions for $k$ in the regions
$\Cin$ and $\Cout$, respectively:
\begin{equation}
\label{eq:tinout}
 \tin=\vartheta(|\eta|-\eta_0)\,,\quad
 \tout \>=\> \vartheta(\eta_0-|\eta|)\,,
\end{equation}
with $\eta_0$ the pseudorapidity corresponding to the opening
(half-)angle of the cone,
$$
\eta_0\>=\> -\ln \tan\frac{\theta_0}{2}.
$$
Note that the fact the probing function $u_\out(k)=1$ for
$k\in\Cin$ means that this parton does not contribute to the
measurement.

Finally, for the associated distribution we introduce a double Mellin
transform:
\begin{equation}
\label{eq:Mellin}
\begin{split}
 \Sigma_{2\ng}(Q,V,\Eout) &= \int\frac{d\nu}{2\pi i\,\nu} 
 \frac{d\nuout}{2\pi i\, \nuout} \> e^{\nu V +\nuout \Eout/Q} \> 
 \tilde\Sigma_{2\ng}(Q,\nu,\nuout)\,,\\
 \tilde\Sigma_{2\ng}(Q,\nu,\nuout) &= \sum_n\int \frac{d\sigma_n}{\sigma_\tot} 
 \prod_{i\in \Ctot}\!\!u_{2\ng}(k_i)\,.
\end{split}
\end{equation}
 The new source function is given by the expression
\begin{equation}
\label{eq:u}
  u_{2\ng}(k) = e^{-\nu\,v(k)} \left(\tin + \tout \,e^{-\nuout\,k_t/Q}\right)
= e^{-\nu\,v(k)} \>-\> \tout e^{-\nu\,v(k)}\left(1- e^{-\nuout\,k_t/Q}\right) .
\end{equation}
It corresponds to measuring $V$ everywhere and $\Eout$ only in
$\Cout$.

In the Mellin space all the distributions become exponents of the
so-called {\em radiators}:
\begin{eqnarray}
\label{eq:exp1}
\tilde \Sigma(Q,\nu) &=& e^{-R(Q,\nu)}\,, \\
\label{eq:exp2}
\tilde \Sigma_\out(Q,\nuout) &=& e^{-R_\out(Q,\nuout)}\,, \\
\label{eq:exp3}
\tilde \Sigma_{2\ng}(Q,\nu,\nuout) &=& e^{-R_{2\ng}(Q,\nu,\nuout)}\,.
\end{eqnarray}

\subsection{Global radiator}

For the global distribution the answer is simple and the radiator is
known to be given by the SL-improved Sudakov expression
\begin{equation}
\label{eq:Rglob}
 R(Q,\nu) = 2C_F \int_0^Q \frac{dk^2_t}{k_t^2}
 \frac{\as(k_t^2)}{\pi} \int_0^{\eta_{\max}}
 d\eta \left(1\!-\!e^{-\nu\,v(k)}\right),
\quad \eta_{\max} = \ln \frac{Q\,e^{-\frac{3}{4}}}{k_t}. 
\end{equation}
In the definition of $\eta_{\max}$ we have incorporated a SL
correction to the ``hard'' piece of the quark splitting function
$$
\alpha P(\alpha) = \frac{1+(1-\alpha)^2}{2} = 1 -
\alpha\left(1-\frac\alpha2\right), \quad \alpha= \frac{k_t}{Q}e^\eta.
$$
It is straightforward to verify then that the virtual SL
contribution due to large rapidities, $\alpha\sim 1$, can be embodied
into rescaling of the upper limit of the $\eta$ integration as
follows:
$$ \int_{k_t/Q}^1 {d\alpha} P(\alpha) \Longrightarrow
\int_0^{\eta_{\max}} d\eta. $$
The two-loop radiator \eqref{eq:Rglob} (with the coupling $\as$ in the
physical ``bremsstrahlung'' scheme~\cite{CMW}) has 
single-logarithmic accuracy. This means that the neglected correction
is $\cO{\as^2\ln \nu}$ which translates into $\as^2\log(Q/V)$ (see,
e.g.~\cite{Broad}).

\subsection{Non-global radiator}

The radiators for both {\em out}\/ and {\em associated}\/ non-global
distributions, \eqref{eq:exp2} and \eqref{eq:exp3}, have the structure
\begin{equation}
\label{eq:Rsplit}
  R_A = R^{(1)}_A + R^{(c)}_A,\qquad A=\out,\, 2\ng\,.
\end{equation}
Here $R_A^{(1)}$ is the standard Sudakov one-gluon emission
contribution and $R_A^{(c)}$ is the term due to correlated multi-gluon
radiation. (The latter is absent (negligible within the SL accuracy)
in the case of a global observable.)

\paragraph{Sudakov piece.}
The Sudakov piece $R_A^{(1)}$,
\begin{equation}
\label{eq:R1}
\begin{split}
 R_A^{(1)} &= 2C_F \int\frac{d^2k_t}{\pi k_t^2}
 \frac{\as(k_t^2)}{\pi} \int_{0}^{\eta_{\max}} d\eta \left[\,1-u_A(k)\,\right],
\end{split}
\end{equation}
is given by the expression \eqref{eq:Rglob} with proper probing
functions:
\begin{subequations}
\label{eq:2ngprob}
\begin{eqnarray}
  \label{eq:2ngproba}
\left[\, 1-u_\out(k)\,\right]   &=& \tout \left(1-e^{-\nuout k_t/Q}\right), \\
\label{eq:2ngprobb}
\left[\, 1-u_{2\ng}(k)\,\right] &=& [\,1-u(k)\,] 
 \>+\> \tout e^{-\nu v(k)}\left(1-e^{-\nuout k_t/Q}\right)
\nonumber\\
&=&  [\,1-u(k)\,] \>+\> e^{-\nu v(k)}\left[\, 1-u_\out(k)\,\right].
\end{eqnarray}
\end{subequations}
We compute first $R^{(1)}_\out$. Substituting \eqref{eq:2ngproba} into
\eqref{eq:R1} we obtain
\begin{equation}
\label{eq:Rout}
  R^{(1)}_\out(Q,\nuout) = 2C_F \int_0^Q \frac{dk^2_t}{k_t^2}
  \frac{\as(k_t^2)}{\pi} \int_0^{\eta_0} d\eta
  \left[1\!-\!e^{-\nuout\,k_t/Q}\right],
\end{equation}
with the rapidity integral restricted to the $\Cout$ region.  Here we
have used that $\eta_0\sim 1 < \eta_{\max}$.  This expression does not
have a collinear singularity and therefore is a SL function.
Therefore we can approximate $[1-u_\out]$ in \eqref{eq:Rout} by a
theta-function $\vartheta(k_t-Q/\nuout)$ and evaluate the inverse
Mellin integral by simply substituting $Q/\Eout$ for $\nuout$.  We
arrive at
\begin{equation}
\label{eq:Routfin}
  R^{(1)}_\out(Q,\nuout) \>\Longrightarrow\> \frac{4C_F}{C_A}\,
  \eta_0\cdot \Delta(Q,\Eout)
\>\equiv\> \cR^{(1)}_\out(Q,\Eout),
\end{equation}
where we have introduced a convenient SL function
\begin{equation}
\label{eq:Ddef}
  \Delta(Q,E) = C_A\int_E^Q \frac{dk_t}{k_t} \frac{\as(k_t^2)}{\pi}.
\end{equation}
Similarly, using \eqref{eq:2ngprobb} in \eqref{eq:R1} gives the
expression for the Sudakov piece of the associated distribution
radiator:
\begin{equation}
\label{eq:R2ngsplit}
  R^{(1)}_{2\ng}(Q,\nu,\nuout) \>=\> R^{(1)}(Q,\nu) \>+\> \delta
  R^{(1)}(Q,\nu,\nuout).
\end{equation}
Here the first term reconstructs the global distribution $\Sigma(Q,V)$
in \eqref{eq:result0} and the addition contribution $\delta R^{(1)}$
to the Sudakov radiator reads
\begin{equation}
\label{eq:R1del}
 \delta R^{(1)}_{2\ng}(Q,\nu,\nuout)= 2C_F \int^Q\frac{dk^2_t}{k_t^2}
 \frac{\as(k_t^2)}{\pi} \int_{0}^{\eta_0} d\eta\,e^{-\nu\,v(k)}
 \left(1- e^{-\nuout\,k_t/Q}\right).
\end{equation}
This expression is identical to that for the out-case,
\eqref{eq:Rout}, apart from the additional factor $\exp(-\nu\,v(k))$.
Since \eqref{eq:R1del} is a subleading SL correction, we can treat
this factor, within our accuracy, as simply proving an additional
restriction upon the transverse momentum. As long as $\eta=\cO{1}$, we
have $v(k)\sim k_t$ and thus
$$
  e^{-v(k)\nu} \quad \Longrightarrow\quad \vartheta(Q-\nu k_t).
$$ 
In $V$-space this translates into the condition
$$
k_t \>\lesssim \> VQ \ll Q.
$$
Then the integral for $\delta R^{(1)}$ becomes the same as that for
$R^{(1)}_\out$ with $Q$ replaced by~$VQ$:
\begin{equation}
\label{eq:R2ngfin}
  \delta R_{2\ng}^{(1)}(Q,\nu,\nuout) 
  \>\Longrightarrow\>  \frac{4C_F}{C_A}\, \eta_0\cdot \Delta(VQ,\Eout)
\>=\> \cR^{(1)}_\out(VQ,\Eout).
\end{equation}
We conclude at the level of the Sudakov contributions to the radiators
the factorized answer \eqref{eq:result0} holds.

Now we have to analyse the specific correlation contribution $R^{(c)}$
in \eqref{eq:Rsplit}.  This is also a SL function whose PT expansion
starts at the level of $(\as\log)^2$.

\paragraph{2-Loop correlation.}

At two loops  $R^{(c)}$  is given by the integral
\begin{equation}
  \label{eq:2loop}
  R_A^{(c)}\>=\>
  \int d\om_2(k_1,k_2)\>[\,U_A(k_1,k_2)-u_A(k_1)u_A(k_2)\,]\>,
\end{equation}
where $d\om_2$ is the distribution for the correlated emission of two
secondary partons off the primary $\qq$ system.
Here $U(k)$ is an ``inclusive'' source function for the parent massive
gluon. As was shown in~\cite{Lucenti}, to properly reconstruct the
two-loop Sudakov expression with the running physical coupling, this
source has to be defined in terms of the single parton source function
$u_A(k) = u_A(k_t,\eta)$
as 
\begin{equation}
  \label{eq:Udef} 
  U_A(k_1,k_2)\>=\> u_A(\sqrt{k_t^2+m^2}, \eta); \quad
  {k}={k}_{1} + {k}_{2},\>\> m^2= k^2.
\end{equation}
Here $\eta$ is the {\em rapidity}\/ of the parent,
$$
{\sqrt{k_t^2+m^2}}e^{\pm\eta} \>=\> 
{k_{t1}}e^{\pm\eta_1} \,+\, {k_{t2}}e^{\pm\eta_2}. 
$$

The distribution $d\om_2$ is singular at $m^2=0$ when the two partons
become collinear or one of them is much softer than the other.
In particular, in the collinear limit the secondary partons obviously
belong to the same angular region ($\Cin$ or $\Cout$) and then, as can
be easily seen from \eqref{eq:2loop}, the combination of the sources
in the square brackets vanishes thus regularizing the singularity. The
same is true for the soft (energy ordered) two-gluon configuration,
provided the gluons are in the same angular region.
This results in a negligible subleading contribution $\cO{\as^2\log}$.

As we will see shortly, the only relevant logarithmically enhanced
contribution emerges in the case when the two secondary partons belong
to the complimentary angular regions, $\Cin$ and $\Cout$, so that the
sources $u(k_1)$ and $u(k_2)$ become different and the cancellation
gets broken.  This is the configuration that gives rise to the
non-global SL corrections, as was found by Dasgupta and
Salam~\cite{DS}.

Given that there are no collinear logarithms, we can simplify
\eqref{eq:2loop} by taking strongly ordered parton energies,
\begin{equation}
\label{eq:enord}
 k_{t2}\ll k_{t1}\ll Q\,.
\end{equation}
In this configuration the correlated 2-gluon distribution 
$d\omega_2$ in \eqref{eq:2loop}
reads~\cite{DMO}
\begin{equation}
  \label{eq:dom2}
  d\om_2(k_1,k_2)\>=\> 2C_F\,C_A\left(\frac{\as}{\pi}\right)^2 \prod_{i=1}^2
  \left(\frac{d\omega_i}{\omega_i}\frac{d\Omega_i}{4\pi}\right) 
  \cdot C(\Omega_1,\Omega_2),  
\end{equation}
The function $C$ depends on parton angles and is given by the expression
\begin{equation}
  \label{eq:Cdef}
\begin{split}
  C(n_1,n_2) =& \frac{(n\bar{n})}{(n n_1)(n_1n_2)(n_2\bar{n})} +
  \frac{(n\bar{n})}{(n n_2)(n_2n_1)(n_1\bar{n})} -
  \frac{(n\bar{n})}{(n n_1)(n_1\bar{n})}\frac{(n\bar{n})}{(n
  n_2)(n_2\bar{n})} \\ & = w_{n\bar{n}}(n_1) \left[\,w_{nn_1}(n_2)+
  w_{n_1\bar{n}}(n_2) -w_{n\bar{n}}(n_2) \,\right], \end{split}
\end{equation}
where 
$$
w_{ab}(c)\equiv \frac{(ab)}{(ac)(cb)},\quad (n_an_b) \equiv
1-\cos\Theta_{ab}
$$
and $n$, $\bar{n}$ stand for the direction 4-vectors of the primary
partons and $n_i$, $i=1,2$ of the secondary gluons.  The angular
distribution \eqref{eq:Cdef} becomes singular when the gluon momenta
$\vec{k}_1$ and $\vec{k}_2$ are parallel.

First we observe that in the soft limit \eqref{eq:2loop} we can put
$m^2=0$ in the source $U$ to approximate
\begin{equation}
\label{eq:uprod}
   [\,U_A(k_1,k_2)-u_A(k_1)u_A(k_2)\,] \>\simeq\> u_A(k_1)\,
   \left[\,1-u_A(k_2)\,\right].
\end{equation}
Let us first analyse the more complicated case of the associated
distribution.  Using the explicit expression for the source
\eqref{eq:u} we obtain two potential contributions:
\begin{subequations} 
\label{eq:gab}
\begin{eqnarray}
\label{eq:good}
&&\eqalign{
k_1 & \in  \Cin,\> k_2\in\Cout: \cr    
&e^{-\nu v(k_1)}\left[1-e^{-\nu v(k_2)- \nuout k_{2t}} \right]
\Longrightarrow  \min\{\Eout,VQ\} < k_{2t} < k_{t1} < VQ;  
} \qquad { } \\
\label{eq:bad}
&&\eqalign{
k_1 & \in \Cout,\> k_2\in\Cin: \cr
&e^{-\nu v(k_1)- \nuout k_{t1}}\left[1-e^{-\nu v(k_2)} \right]
\Longrightarrow  VQ < k_{2t} < k_{t1} < \min\{\Eout,VQ\}.
}
\end{eqnarray}
\end{subequations}
We see that the only contributing region is that in \eqref{eq:good}
where a harder gluon is close to the jet axis while a softer one
contributes to the out-of-jet $k_t$--flow, provided $\Eout\ll VQ$.
Therefore in \eqref{eq:2loop} we can substitute
\begin{equation}
\label{eq:ulead}
[U_{2\ng}(k_1,k_2)-u_{2\ng}(k_1)u_{2\ng}(k_2)] \>\Rightarrow \>
\tin(k_1)\tout(k_2)\, \vartheta(VQ-k_{1t})\,\vartheta(k_{2t}-\Eout)
\end{equation}
and we get at two loops
\begin{equation}
\label{eq:Rc1}
 R^{(c)}_{2\ng} \>\Longrightarrow\> \frac{C_F}{C_A}
\>\Delta^2(VQ,\Eout)\cdot F(\eta_0).
\end{equation}
Here $F(\eta_0)$ is given by the angular integral 
\begin{equation}
\label{eq:F}
 F(\eta_0)= \int_{\Cin}\!\frac{d^2\Om_1}{4\pi}
 \int_{\Cout}\!\frac{d^2\Om_2}{4\pi}\, C(n_1,n_2)
= \frac{\pi^2}{6}-{\rm Li}_2
\left(e^{-4\eta_0}\right) .
\end{equation}

The same conclusion holds for the out-distribution:
\begin{equation}
\label{eq:uoutlead}
[U_{\out}(k_1,k_2)-u_{\out}(k_1)u_{\out}(k_2)] \>\Rightarrow \>
\tin(k_1)\tout(k_2)\, \vartheta(Q-k_{1t})\,\vartheta(k_{2t}-\Eout).
\end{equation}
Comparing this with \eqref{eq:ulead} we
see that this case reduces to setting $V=1$ in \eqref{eq:gab}, and we obtain
\begin{equation}
\label{eq:Rcout}
 R^{(c)}_{\out} \>\Longrightarrow\> \frac{C_F}{C_A}
\>\Delta^2(Q,\Eout)\cdot F(\eta_0).
\end{equation}
The result \eqref{eq:Rc1} is just the same expression that we derived
for the away from jet distribution, except that what was $Q$ there is
now replaced by $VQ$.

This completes the proof of \eqref{eq:result0} with account of the
2-loop gluon correlations.

So we can write the integrated non-global distributions \eqref{eq:2distr} 
with the SL accuracy as 
\begin{subequations}
\label{eq:result1}
\begin{eqnarray}
\label{eq:result1out}
\Sigma_{\out}(Q,\Eout) &=& e^{-\cR^{(1)}_\out(Q,\Eout)}\,
e^{-\cR^{(c)}_{\out}(Q,\Eout)} ,\\
\label{eq:result12ng}
\Sigma_{2\ng}(Q,V,\Eout) &=&
 e^{-\cR^{(1)}_\out(VQ,\Eout)}\,e^{-\cR^{(c)}_{2\ng}(Q,V,\Eout)} \cdot
 \Sigma(Q,V),
\end{eqnarray}
\end{subequations}
where $\cR^{(1)}_\out$ was defined in \eqref{eq:Routfin}.  We have
shown in this section that with account of the 2-gluon correlation
$(\cO{\Delta^2})$
\begin{equation}
\label{eq:uguali}
 \cR^{(c)}_{2\ng}(Q,V,\Eout) \>=\> \cR^{(c)}_{\out}(VQ,\Eout).
\end{equation}
In the following section we show that \eqref{eq:uguali} hold at all
orders in $\Delta$.

\section{\label{sec4}Multi-gluon correlations in all orders}

Since there is no collinear singularities in $R^{(c)}$, all SL
contributions arise from ensembles of soft gluons with strongly
ordered energies -- the leading logarithmic soft approximation.  As a
result of the strong energy ordering, it is only the {\em hardest}\/
of these soft gluons in the angular region $\Cout$ that contributes to
$\Eout$ (since the contribution of {\em softer}\/ gluons are
negligible in the leading soft approximation).

Therefore, to collect these specific SL contributions to the
non-global distributions (both the {\em out}- and {\em associated}\/
distributions) it suffices to analyse multi-gluon systems with many
gluons in $\Cin$ and a single one in $\Cout$. Let us denote the
momentum of this gluon by $q$.
Then, among the gluons $k_i\in\Cin$ it suffices to consider only those
that are {\em harder}\/ then $q$, since only such gluons may affect
the radiation of $q$. Softer ones, $k_i\ll q$, don't contribute to the
measurement due to real-virtual cancellations (except as
power-suppressed corrections).

With account of multi-gluon effects the non-global correlation
function $\cR^{(c)}_A$ becomes much more involved. At the same time,
the structure of its dependence on $Q$ and $V$ remains the same as in
the 2-loop case.  Indeed, the logarithmic integration in the energy of
the hardest gluon $k_1$ in the $\Cin$ region is limited from above
either by a pure phase space restriction $k_{10}<Q$ for the case of
$\cR^{(c)}_\out$ or, alternatively, by $k_{10}<VQ$ in the case of the
associated distribution $\cR^{(c)}_{2\ng}$. Since this upper limit is
the only place where the dependence on $Q$ ($QV$) enters, we
automatically derive the relation~\eqref{eq:uguali}.  Therefore it
suffices to consider $\cR^{(c)}_\out(Q,\Eout)$.  Hereafter we suppress
the subscript and study
$$ 
 \cR^{(c)}(Q,\Eout) \>\equiv
 \>\cR^{(c)}_\out(Q,\Eout) \>=\> \cR^{(c)}_{2\ng}(Q,V=1,\Eout).  
$$

\paragraph{Generalisation.}
For the sake of generalisation the 2-loop result can be rephrased as
follows.  We take $k\in\Cin$ and $q\in\Cout$, $q\ll k$, and write
($\bas\equiv C_A\as/\pi$)
\begin{equation}
\begin{split}
  \cR^{(c)}(Q,\Eout) =& 
 \int_{\Eout}^{Q}\frac{dk_0}{k_0}\int_{\Cin} \frac{d\Omega_k}{4\pi} \>
 \bas\, w_{n\bar{n}}(n_k) \\ 
 & \times \left\{ r_{nn_k}(k_0,\Eout) + r_{n_k\bar{n}}(k_0,\Eout) -
 r_{n\bar{n}}(k_0,\Eout) \right\}   ,
\end{split}
\end{equation}
where $r_{ab}$ is the 1-loop radiator for the out-distribution
generated by a colorless dipole $ab$ with $a,b\in\Cin$:
\begin{equation}
 r_{ab}(Q,\Eout) = \frac{2C_F}{C_A}
 \int_{\Eout}^{Q} \frac{dq_0}{q_0} \int_{\Cout} \frac{d\Omega_q}{4\pi}\>
\bas\, w_{ab}(n_q)\,.
\end{equation}
In the large-$N_c$ limit the structure of soft multi-gluon
multiplication can be described in terms of the iterative procedure
(see \cite{BCM}):
\begin{equation}
 W_{ab}(1,\ldots ,m) = w_{ab}(\ell) \, W_{a\ell}(1,\ldots ,\ell\!-\!1)
 W_{\ell b}(\ell\!+\!1, \ldots ,m), \quad W_{ab}(1)=w_{ab}(1),
\end{equation}
where $\ell$ is one of $m$ soft gluons emitted by the colour singlet
dipole $(ab)$. We choose $\ell$ to be the hardest of the secondary
gluons.  Following this path we arrive at the following
generalisation:
\begin{subequations}
\label{eq:eqRc}
\begin{equation}
 \cR^{(c)}_{ab}(Q,\Eout) = 
\int\limits_{\Eout}^Q \!\!\frac{dk_0}{k_0}
\int\limits_{\Cin} \!\!\frac{d\Omega_k}{4\pi} 
\> \bas\, w_{ab}(n_k)
\left\{1- \frac{Z_{a n_k}(k_0,\Eout)Z_{n_k b}(k_0,\Eout)}{Z_{ab}(k_0,\Eout)} 
 \right\}\!,\quad { }
\end{equation}
where
\begin{equation}
-\ln Z_{ab}(E,\Eout) \>=\> r_{ab}(E,\Eout) + \cR^{(c)}_{ab}(E,\Eout).
\end{equation}
\end{subequations}
In \eqref{eq:eqRc} $Z_{ab}$ is the generating function describing the
gluon cascade which originates from the $ab$-dipole.  We remark that
the coupled equations \eqref{eq:eqRc} describing the large-angle
(hedgehog) gluon configurations have a highly non-linear structure.

Once the equation is solved, we substitute the directions of the
primary partons $n\bar{n}$ for $ab$ to obtain
\begin{equation}
\label{eq:cRcnn}
\cR^{(c)}_\out(Q,\Eout) \>=\> \cR^{(c)}_{n\bar{n}}(Q,\Eout).
\end{equation}
Recall that to obtain the all-order correlation function
$\cR^{(c)}_{2\ng}(Q,V,\Eout)$ for the associated distribution we
simply have to substitute $Q\to VQ$ in \eqref{eq:cRcnn}.

\paragraph{The asymptotic regime.}

The behaviour of the answer in the asymptotic limit $\as\ln
(Q/\Eout)\gg 1$ ($\as\ln (VQ/\Eout)\gg 1$) was analysed in \cite{BMS}.
For the partons $a$ and $b$ inside the same $\Cin$ cone and
$\theta_{ab}\ll1$
\begin{equation}
\label{eq:largeLn}
\begin{split}
 Z_{ab}\simeq h\left(\frac{\theta^2_{ab}}{2\theta_c^2}\right)\,,
 \quad \theta_c\simeq \lam(\eta_0)e^{-\frac{c}{2}\De}\,, 
 \quad c\simeq 2.5
\end{split}
\end{equation}
with $c$ a universal number and $\lam$ a finite factor which depends
on the opening angle.  Here $\theta_c$ is a ``critical angle'' which
depends on the SL parameter $\Delta$ defined in \eqref{eq:Ddef},
and $h(x)$ is a universal function which is normalized, $h(0)=1$, and
decreases fast for large $x$ as
$$
  h(x) \propto \exp(-\frac{c}{2}\ln^2x).  
$$
When $\Delta$ is large, $Z$ can be approximated by a theta-function
$$ 
 Z_{ab}(V,\Eout) \simeq \vartheta\left(\theta_c(\Delta) -
 \theta_{ab}\right).
$$
The dependence on $\theta_0$ 
becomes subleading in $\Delta$, and we get an asymptotic estimate
\begin{equation}
\label{eq:asympt}
  \cR^{(c)}(V,\Eout) \>=\>  c\cdot \De^2 \> \left( 1+\cO{\Delta^{-1}}\right).
\end{equation}
Let us remark that the $\cR^{(1)}_{\out}$ (Sudakov) contribution to
$\ln\Sigma$ is in this limit comparable with the subleading term in
\eqref{eq:asympt} which is geometry-dependent and could be computed by
a numerical solution of the equation \eqref{eq:eqRc} or by the methods
of~\cite{DS,DScase}.

\section{Conclusions}

We considered the associated measurement of a global event shape
variable $V$ (e.g.\ thrust) and the away from jets transverse energy
flow $\Eout$ (the ``flow/shape'' correlation~\cite{BKS}) in $\ee$
annihilation.  We showed that the corresponding integrated
distribution $\Sigma_{2\ng}$ defined in \eqref{eq:ng} factorizes into
the product of the global distribution in $V$ \eqref{eq:tot} and the
(non-global) SL distribution in $\Eout$ \eqref{eq:out} taken at the
rescaled energy $VQ$, as stated in \eqref{eq:result0}.  This result
holds with DL$+$SL accuracy, including resummation of the SL terms in
the ratio of the two characteristic scales, $(\as\ln(VQ/\Eout))^n$.

This result hold also for other shape variables, in particular, for
the broadening $B$, in which case the analysis of the global spectrum
is more involved~\cite{Broad}.

In the most interesting limit $VQ\gg \Eout$ the distribution
$\Sigma_{2ng}$ describes also the case when the $V$ observable is
measured only inside $\Cin$ (rather than in the whole phase space), as
has been suggested in~\cite{BKS}.

\acknowledgments
We wish to thank Gavin Salam for valuable discussions.

\end{document}